\documentclass[preprint,showpacs,preprintnumbers,amsmath,amssymb,showkeys]{revtex4}

\usepackage{graphicx}
\usepackage{dcolumn}

\usepackage{bm}

\begin{document}

\title{Non-vanishing $U_{e3}$ under $S_3$ symmetry}


\author{Kim Siyeon} \email{siyeon@cau.ac.kr}
\affiliation{Department of Physics,
        Chung-Ang University, Seoul 156-756, Korea}

\date{June 12, 2012}
\begin{abstract}
This work proposes two models of neutrino masses that predict non-zero $\theta_{13}$ under the non-Abelian discrete flavor symmetry $\mathbb{S}_3\otimes\mathbb{Z}_2$. We advocate that the size of $\theta_{13}$ is understood as a group theoretical consequence rather than a perturbed effect from the tri-bi-maximal mixing. So, the difference of two models is designed only in terms of the flavor symmetry, by changing the charge assignment of righthanded neutrinos. The PMNS matrix in the first model is obtained from both mass matrices, charged leptons giving rise to non-zero $\theta^l_{13}$ and neutrino masses giving rise to tri-bi-maximal mixing. The physical mixing angles are expressed by a simple relation between $\theta^l_{13}$ and tri-bi-maximal angles to fit the recent experimental results. The other model generates PMNS matrix with non-zero $\theta_{13}$, only from the neutrino mass transformation. The 5 dimensional effective theory of Majorana neutrinos obtained in this framework is tested with phenomenological bounds in the parametric spaces $\sin\theta_{23}, \sin\theta_{12}$ and $m_2/m_3$ vs. $\sin\theta_{13}$.
\end{abstract}


\maketitle \thispagestyle{empty}


\section{Introduction}

Recent long-baseline neutrino experiments, T2K \cite{Abe:2011sj} and MINOS \cite{Adamson:2011qu}, gave rise to the first indications to a non-zero $U_{e3}$. Following reactor neutrino experiments successfully presented certain values of $\sin^22\theta_{13}$. Double Chooz reported their first result $\sin^2(2\theta_{13})=0.085$ at the 68\% CL \cite{Abe:2011fz}. Daya Bay narrowed down the range to $\sin^2(2\theta_{13})=0.092 \pm 0.005(\mathrm{syst.})$ at 5.2$\sigma$ \cite{An:2012eh}, and RENO reported a definitive result with a value of $\sin^2(2\theta_{13})=0.113 \pm 0.019(\mathrm{syst.})$ at 4.9$\sigma$ \cite{Ahn:2012nd}.
The current bound on other angles, determined from neutrino oscillation experiments, are $0.490\leq \sin\theta_{12}\leq0.632,$ and $0.583\leq \sin\theta_{23}\leq0.825$ at the $90\%$ CL. The current data also include the mass squared differences that are accompanied by solar and atmospheric neutrino oscillations, $\Delta m_{sol}^2 \simeq (7^{+10}_{-2})\times 10^{-5}eV^2$ and $\Delta m_{atm}^2 \simeq (2.5^{+1.4}_{-0.9})\times 10^{-3} eV^2$, respectively \cite{Nakamura:2010zzi,GonzalezGarcia:2007ib}.

Non-Abelian discrete symmetries have provided theoretical frameworks for neutrino masses with tri-bi-maximal(TBM) mixing \cite{Harrison:2002er,Harrison:2003aw} with $\sin\theta_{12}=1/\sqrt{3}, ~ \sin\theta_{23}=1/\sqrt{2},$ and $\sin\theta_{13}=0$ \cite{Zee:2005ut,Haba:2006dz,Lam:2008rs,Hagedorn:2006ug,Zhang:2006fv,Bazzocchi:2009pv,Yang:2011fh,Morisi:2011ge,Park:2011zt}. Due to the signals from recent measurements of $\theta_{13}$, its non-zero value, which is still small relative to other two angles, is considered as being generated by a mechanism based on the symmetrical background rather than being a perturbation effect.

Two models with non-zero $U_{e3}$ are introduced using $\mathbb{S}_3\otimes\mathbb{Z}_2$ flavor symmetry. Both models have the same field contents with the same flavor charges, except for two righthanded neutrinos. Whether the $\mathbb{S}_3$ representations of the two fields are double $\mathbf{1'}$s or a single $\mathbf{2}$, the non-zero $U_{e3}$ is obtained in charged lepton masses or in neutrino masses. Besides the Standard Model(SM) Higgs, a few scalar multiplets are added. The $\mathbb{Z}_2$ which commutes with $\mathbb{S}_3$ divides the fields by their parity, in the sense that all SM fields have even parity and so their couplings are not affected by the $\mathbb{Z}_2$ symmetry. The $\mathbb{Z}_2$-odd righthanded neutrinos couple with only $\mathbb{Z}_2$-odd scalar fields to make 5-dimensional Majorana masses in an effective theory. Using a simple assumption of Yukawa coupling constants, predictions of the mass ratios and mixing angles are presented.

This paper is organized as follows: Section II introduces the representations of flavor symmetry $\mathbb{S}_3$, and contains the construction of Yukawa interactions of SM charged leptons with $\mathbb{Z}_2$-even Higgs contents. In Section III, two models with non-zero $U_{e3}$ are presented. The first model obtains the PMNS angle by a simple relation of TBM angles and the mixing angle $\theta_l$ of charged leptons. In the second model, the transformation of neutrino mass matrix becomes the PMNS matrix to the leading order. The predictions are examined in the figures. The conclusion section contains a summary and mentions some exclusion regions as the prediction, and an appendix is attached to describe the interactions of Higgs fields and their vevs to make the potential minimum.

\section{Discrete flavor symmetry $\mathbb{S}_3$ and Yukawa interaction}

The minimal non-Abelian discrete symmetry $\mathbb{S}_3$ is the group of the permutation of the three sides of an equilateral triangle. There are six elements of the group in three classes, and their irreducible representations are $\mathbf{1},~ \mathbf{1'},$ and $\mathbf{2}.$ Its character table is mentioned in many models \cite{Morisi:2011pm,Meloni:2010aw,Dong:2011vb,Chu:2011jg}.

The Clebsch-Gordon coefficients in the real representations are given by the following product rules \cite{Ma:2004pt,Chen:2004rr},
    \begin{eqnarray}
    && \mathbf{1'}\times\mathbf{1'} = \mathbf{1}~:~ ab, \label{prod1}\\
    && \mathbf{1'}\times\mathbf{2} = \mathbf{2}~:~
        \left(\begin{array}{r}ab_2\\-ab_1 \end{array}\right)\\
    && \mathbf{2}\times\mathbf{2} =
        \mathbf{1}+\mathbf{1'}+\mathbf{2}, \label{prod3} \\
    && \begin{array}{lcl}
        \mathbf{1} & : & (a_1b_1+a_2b_2) \\
        \mathbf{1'} & : & (a_1b_2-a_2b_1) \\
        \mathbf{2} & : &
        \left(\begin{array}{l}a_2b_2-a_1b_1\\a_1b_2+a_2b_1 \end{array}\right), \nonumber
        \end{array} \label{prod5}
    \end{eqnarray}

    \begin{table}
    \caption{Group representation of SM particles}
    \begin{ruledtabular}
    \begin{tabular}{c|ccc|c}
        $\mathrm{Rep.}$  & $(\mathbf{1'},1)_F$ & $(\mathbf{2},1)_F$ &  & $(\mathbf{1},1)_F$ \\ \hline
        $(\mathbf{2},-1/2)_G$  & $l_e$ & $L_\alpha : (l_\mu, l_\tau)$ & & $H$ \\
        $(\mathbf{1},-1)_G$  & $e_r$ & $R_\alpha : (\mu_r, \tau_r)$ & & ... \\
    \end{tabular}
    \end{ruledtabular}
    \end{table}

Here, an Abelian discrete symmetry $\mathbb{Z}_2$ is also adopted, which is the parity that distinguishes extra particles from the SM contents.
The SM fields are assigned to representations of $\mathbb{S}_3\otimes\mathbb{Z}_2$ as listed in Table I, where the SU(2) representation and hypercharge of a field are denoted by the subscription `$G$' of the gauge symmetry, and the $\mathbb{S}_3$ representation and  $\mathbb{Z}_2$ charge of the field are denoted by the subscription `$F$'. Although the $\mathbb{Z}_2$ charges of fields are presented in Table I, $\mathbb{Z}_2$ symmetry does not affect the interactions among $\mathbb{Z}_2$-even SM fields. Then, the Lagrangian of Yukawa couplings of the charged leptons and the Higgs scalar doublet $H$ is
    \begin{eqnarray}
    -\mathcal{L}_{SM}=c_1 H \overline{e}_r l_e + c_2 H \overline{R}_\alpha L_\alpha, \label{sm}
    \end{eqnarray}
where the SU(2) fermion doublet $l_e$ is a flavor singlet but $l_\mu$ and $l_\tau $ belong to a doublet such as $L_\alpha\equiv(l_\mu, l_\tau)$ under $\mathbb{S}_3$. Also,  the righthanded charged lepton singlet $e_r$ is a flavor singlet while $\mu_r$ and $\tau_r$ belong to a doublet such as $R_\alpha\equiv(\mu_r, \tau_r)$. The Higgs scalar doublet $H$ of SM is involved in the above interactions as a flavor singlet.
The Higgs self potential is
    \begin{eqnarray}
    V_H = m_H^2H^\dagger H + \frac{1}{2}\eta(H^\dagger H)^2,
    \end{eqnarray}
so that, after spontaneous SU(2) symmetry breaking, three Dirac mass matrices of the charged leptons from the Yukawa couplings become
    \begin{eqnarray}
        c_i \langle H \rangle \sim \left(
        \begin{array}{ccc}
        c_1 v & 0 & 0 \\
        0 & c_2 v & 0 \\
        0 & 0 & c_2 v
        \end{array} \right).\label{smLepton}
    \end{eqnarray}
It is shown that for the flavor model we can build a basis where the matrix of charged lepton masses is diagonal and $m_2=m_3$, if only SM fields contribute to generate the Dirac masses. It follows that the right mass hierarchy is obtained from additional Yukawa couplings with additional scalar fields beyond the SM.

There is an additional Higgs scalar particle that couple with SM leptons, which is represented by, under $\mathbb{S}_3\otimes\mathbb{Z}_2$,
    \begin{eqnarray}
        (\mathbf{2},1)_F & : & \Phi~(\varphi_1, \varphi_2).
    \end{eqnarray}
The interactions of only $\mathbb{Z}_2$-even Higgs particles, $H,$ and $\Phi$, among themselves are
    \begin{eqnarray}\begin{array}{ccl}
    V_e(H,\Phi)   & = & V_H + m_\varphi^2 \Phi^\dagger\Phi +
            \frac{1}{2}\Lambda(\Phi^\dagger\Phi)_r^2  \\
            & + & \lambda(\Phi^\dagger\Phi)_1(H^\dagger H)_1 +
            \lambda'(\Phi^\dagger H)_2(H^\dagger\Phi)_2 \\
            & + & \lambda''\{(\Phi^\dagger H)_2^2 + h.c.\} \\
            & + & \kappa\{(\Phi^\dagger\Phi)_2(\Phi^\dagger H)_2 +
            h.c.\},
    \label{potentialNH}
    \end{array}\end{eqnarray}
where the term $\frac{1}{2}\Lambda(\Phi^\dagger\Phi)_r^2$ include such three contributions as
    \begin{eqnarray}
    \Lambda (\Phi^\dagger\Phi)^2 = \lambda_a(\Phi^\dagger\Phi)_1^2 + \lambda_b(\Phi^\dagger\Phi)_{1'}^2 + \lambda_c(\Phi^\dagger\Phi)_2^2,
    \label{3reps}
    \end{eqnarray}
since the product $\Phi^\dagger\Phi$ can be any of the following representations,
$(\mathbf{1},1), ~(\mathbf{1'},1)$ or $(\mathbf{2},1)$ of $\mathbb{S}_3\otimes\mathbb{Z}_2$.
According to the processes in Appendix in order to make the potential minimum, the vevs are obtained such that $\langle H\rangle=\langle H^\dagger\rangle=v$,  $\langle\varphi_1\rangle=v_1$, and $\langle\varphi_2\rangle=0$.

If the masses of charged leptons were obtained by using the Yukawa couplings in Eq.(\ref{smLepton}), the muon and tau lepton could have the same mass, $m_\mu=m_\tau$. Here, we introduce an additional contribution to the masses derived from the Yukawa couplings with another Higgs $\Phi$.
    \begin{eqnarray}
    -\Delta\mathcal{L}_{SM}=c_3 \Phi \overline{R}_\alpha L_\alpha + c_4 \Phi \overline{e}_r L_\alpha + c_5 \Phi \overline{R}_\alpha l_e, \label{sm_extended}
    \end{eqnarray}
while the Yukawa couplings of quarks are protected from the non-SM additional Higgs $\Phi$, since the flavor symmetry is leptonic so that all quarks are $\mathbb{S}_3$ singlets and $\mathbb{Z}_2$-even.
The Dirac mass matrix of charged leptons derived from both Eq.(\ref{sm}) and Eq.(\ref{sm_extended}) has the following form
    \begin{eqnarray}
        M_{l} \sim \left(
        \begin{array}{ccc}
        c_1 v & 0 & c_4 v_1 \\
        0 & c_2 v - c_3 v_1 & 0 \\
        c_5 v_1 & 0 & c_2 v + c_3 v_1
        \end{array} \right).\label{MassLepton}
    \end{eqnarray}
The masses of leptons are obtained from $U_l M_l^\dagger M_l U_l^\dagger = Diag(m_e^2, m_\mu^2, m_\tau^2)$. The transformation matrix $U_l$ is required for Pontecorvo-Maki-Nakagawa-Sakata(PMNS) matrix along with the transformation of neutrino masses $U_\nu$ such that $U_{PMNS}=U_l^\dagger U_\nu$. We denote the 1-3 block of the matrix $M_l^\dagger M_l$ by $K$ such as
    \begin{eqnarray}
    K \equiv \left(
        \begin{array}{cc}
        |c_1|^2 v^2 + |c_5|^2 v_1^2 & c_1^*c_4 v v_1 + c_5^* v_1 (c_2 v + c_3 v_1) \\
        c_v^*c_1 v v_1 + (c_2^* v + c_3^* v_1)c_5 v_1  & |c_2 v + c_3 v_1|^2
        \end{array} \right), \label{mdaggerm}
    \end{eqnarray}
which is plausible by the relation in terms of the masses and the mixing angle as in $K=R(\theta_l, \delta_l)Diag(m_e^2, m_\tau^2)R^\dagger(\theta_l, \delta_l)$, where the 1-3 block rotation $R_{13}(\theta_l, \delta_l)$ is given by
    \begin{eqnarray}
    R(\theta_l, \delta_l) \equiv \left(
        \begin{array}{cc}
        \cos\theta_l & \sin\theta_l e^{-i\delta_l} \\
        -\sin\theta_l e^{i\delta_l} & \cos\theta_l
        \end{array} \right).\label{TansLepton}
    \end{eqnarray}
The elements of the matrix $K$ in Eq.(\ref{mdaggerm}) are described by physical parameters,
    \begin{eqnarray}
    K_{11} &=& m_e^2\cos^2\theta_l + m_\tau^2\sin^2\theta_l, \nonumber \\
    K_{22} &=& m_\tau^2\cos^2\theta_l + m_e^2\sin^2\theta_l, \label{elementK}\\
    K_{12} &=& K_{21}^* = \left( m_\tau^2e^{i\delta_l} - m_e^2e^{-i\delta_l}\right)\cos\theta_l\sin\theta_l. \nonumber
    \end{eqnarray}
In opposite way, the mixing angle $\theta_l$ and the phase $\delta_l$ are obtained from the elements in Eq.(\ref{elementK}) as
    \begin{eqnarray}
    \tan2\theta_l \cos\delta_l= \frac{K_{12}+K_{12}^*}{K_{22}-K_{11}},\label{thetadelta}
    \end{eqnarray}
or from the matrix in Eq.(\ref{mdaggerm})
    \begin{eqnarray}
    \tan2\theta_l = \frac{2 \mathrm{Re}[c_1^*c_4 v v_1 + c_5^* v_1 (c_2 v + c_3 v_1)]}{|c_2 v + c_3 v_1|^2-|c_1|^2 v^2 - |c_5|^2 v_1^2}.\label{theta}
    \end{eqnarray}
In general, the squared masses can be expressed in the following way,
    \begin{eqnarray}
    m_e^2 &=& \frac{1}{2}\left( K_{22} + K_{11} \right)
            - \frac{1}{2}\left( K_{22} - K_{11} \right)\sqrt{1 +\tan^2 2\theta_l\cos^2\delta_l}, \nonumber \\
    m_\tau^2 &=& \frac{1}{2}\left( K_{22} + K_{11} \right)
            + \frac{1}{2}\left( K_{22} - K_{11} \right)\sqrt{1 +\tan^2 2\theta_l\cos^2\delta_l},
    \label{msquare}
    \end{eqnarray}
where $m_\mu^2 = |c_2 v - c_3 v_1|^2$.
For $\tan^2 2\theta_l \ll 1$, the squared masses are approximated to
    \begin{eqnarray}
    m_e^2 &\approx& K_{11} = |c_1|^2 v^2 + |c_5|^2 v_1^2, \nonumber \\
    m_\tau^2 &\approx& K_{22} = |c_2 v + c_3 v_1|^2 , \label{diagmm}
    \end{eqnarray}
where $m_e^2$ and $m_\tau^2$ are derived from the independent Yukawa couplings from each other, and $M_l$ is close to a diagonal matrix due to the suppressed numerator in Eq.(\ref{theta}).
For $\tan^2 2\theta_l \gg 1$, the squared masses in Eq.(\ref{msquare}) reduce to
    \begin{eqnarray}
    m_e^2 &=& \frac{1}{2}\left( K_{22} + K_{11} \right)
            - \frac{1}{2}\left( K_{22} - K_{11} \right)\tan 2\theta_l\cos\delta_l,
            \nonumber \\
    m_\tau^2 &=& \frac{1}{2}\left( K_{22} + K_{11} \right)
            + \frac{1}{2}\left( K_{22} - K_{11} \right)\tan 2\theta_l\cos\delta_l. \label{largetan}
    \end{eqnarray}
The strong hierarchy between $m_e^2$ and $m_\tau^2$ and the large mixing angle require a careful fine tuning in Eq.(\ref{largetan}).

\section{Neutrino Masses under $\mathbb{S}_3$ symmetry}

Two models are presented to explain neutrino masses in the normal hierarchy and the mixing matrix. Both models contain $\mathbb{Z}_2$-odd additional particles, which can be distinguished from $\mathbb{Z}_2$-even SM fields. The righthanded neutrinos are characterized by $\mathbb{Z}_2$ charge -1 and coupled with  $\mathbb{Z}_2$-odd Higgs fields, while the SM righthanded leptons are all assigned to the charge +1 and coupled with  $\mathbb{Z}_2$-even Higgs fields. Those additional Higgs contents are
    \begin{eqnarray}
        (\mathbf{2},-1)_F & : & \Sigma~(\sigma_1, \sigma_2)\nonumber \\
        (\mathbf{1},-1)_F & : & h.
    \end{eqnarray}
The Higgs potential which include the interactions of $h$ and $\Sigma$ is presented in Eq.(\ref{potential_sigma}). Its minimum is obtained by the real vevs of the Higgs fields, which are denoted as $\langle h\rangle=\langle h^\dagger\rangle=u$ and $\langle \sigma_1\rangle= \langle \sigma_2 \rangle=w$.

As far as the mixing of neutrino mass matrix is concerned, one model generates TBM, and the other generates nonzero $\theta_{13}$ in the PMNS matrix. Both types of mixing matrices are derived in terms of the $\mathbb{S}_3$ symmetry and its breaking mechanism by the vacuum expectation value(vev) of an $\mathbb{S}_3$-doublet scalar field. The field contents and their charge assignments in the two models are identical to each other except the flavor charges of the righthanded neutrinos. The operators for Majorana masses have four external lines with an internal line, as shown in Fig.1. If the internal line is a heavy righthanded neutrino which has its ends coupled in a Yukawa interaction, the process giving rise to low energy effective masses is equivalent to the Seesaw Mechanism. Here, we describe the generation of neutrino masses while comparing the two models, the difference between which originated from the choice of group representations for the internal righthanded neutrinos.
\begin{figure}
\resizebox{50mm}{!}{\includegraphics[width=0.75\textwidth]{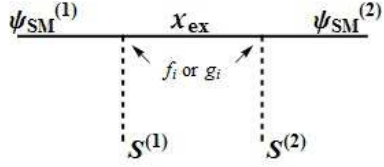}}
\caption{\label{fig:rge_msqr} Effective operator to give a mass to a Majorana particle. A 5-dimensional interaction of two scalars, $S^{(1)}$ and $S^{(2)}$, and two SM fermions, $\psi_{SM}^{(1)}$ and $\psi_{SM}^{(2)}$, is obtained by integrating out the heavy fermion $x_{\mathrm{ex}}$ in the internal line.}
\end{figure}

\subsection{Neutrino model for tri-bi-maximal $U_\nu$ }

All additional fields beyond the SM, including righthanded neutrinos, are distinguished from the SM particles by $\mathbb{Z}_2$ parity. All the SM fields are $\mathbb{Z}_2$ even, so that the parity does not affect any interaction of SM particles. Additional Higgs scalars, $\Sigma \equiv (\sigma_1,\sigma_2)$ and $h$, and right-handed neutrinos, $n_1, n_2, n_3$, all have $\mathbb{Z}_2$-odd quantum number. Their representation under the gauge symmetry is $(\mathbf{1},0)_G$, and their representations under the flavor symmetry are
    \begin{eqnarray}
    \begin{array}{lcl}
        (\mathbf{1},-1)_F & : & n_1 \nonumber \\
        (\mathbf{1'},-1)_F & : & n_2, n_3.
        \end{array}
    \end{eqnarray}
Their Yukawa interactions are as follows:
    \begin{eqnarray}
    -\mathcal{L}_{ext}=f_0 h \overline{n}_1 l_e + f_1 \Sigma \overline{n}_1 L_\alpha + f_2 \Sigma \overline{n}_2 L_\alpha + f_3 \Sigma \overline{n}_3 L_\alpha, \label{YukawaTbm}
    \end{eqnarray}
where $n_3$ is redundant so that the same result can be obtained by just two righthanded neutrinos. But we keep the two identical singlets, $n_2$ and $n_3$, for the comparison with other representation for them in another model.
The couplings of $\Sigma, ~n_i$ and $L_\alpha$ can be rephrased as $f_1 \Sigma \overline{n}_1 L_\alpha = f_1 \overline{n}_1 (\sigma_1 l_\mu + \sigma_2 l_\tau)$, $f_2 \Sigma \overline{n}_2 L_\alpha = f_2 \overline{n}_2 (\sigma_1 l_\tau - \sigma_2 l_\mu)$, and $f_3 \Sigma \overline{n}_3 L_\alpha = f_3 \overline{n}_3 (\sigma_1 l_\tau - \sigma_2 l_\mu)$. The gauge singlets $n_i$ have Majorana masses as in $\frac{1}{2} M_i n_i n_i$, while the $n_2$ and $n_3$ have a cross term such as $M_x n_2 n_3$. If Majorana neutrinos $n_i$ are very heavy, any two Yukawa couplings could be linked to each other as shown in Fig.1.

The vertex $f_i S^{(1)}\psi_{SM}^{(1)}\chi_{ex}$ or $f_i S^{(2)}\psi_{SM}^{(2)}\chi_{ex}$ in Figure 1 can correspond to any term in Eq.(\ref{YukawaTbm}) if the $n_i$ that is substituted into $\chi_{ex}$ is the same for both vertices. Then, an effective Lagrangian is obtained by integrating out the internal heavy fermion $\chi_{ex}$.
    \begin{eqnarray}
    -\mathcal{L}_{eff}
    & = & \frac{f_0^2}{M_1} h h l_e l_e +
        \frac{f_0 f_1}{M_1} ( h\sigma_1 l_e l_\mu + h\sigma_2 l_e l_\tau ) \nonumber \\
    & + & \frac{f_1^2}{M_1} ( \sigma_1 l_\mu + \sigma_2 l_\tau )^2 \\
    & + & (\frac{f_2^2}{M_2} + \frac{f_3^2}{M_3} + 2\frac{f_2 f_3}{M_x}) ( \sigma_2 l_\mu - \sigma_1 l_\tau )^2. \nonumber
    \label{effective1}
    \end{eqnarray}
Here, $M_i$ and $M_x$ are the elements of mass matrix of singlet Majorana neutrinos $n_i$ for $i=1-3$. Since $n_1, n_2,$ and $n_3$ all belong to different representations, their Yukawa coupling constants $f_i$ can be different and so can be their masses, $M_i$. When the scalar fields obtain vevs by spontaneous breaking of $\mathbb{S}_3$ symmetry, the above 5-dimensional interactions reduce to low-energy effective mass terms of light neutrinos $M^{(\nu)}_{ij}\nu_i\nu_j$. The symmetric matrix is
    \begin{eqnarray}
    M^{(\nu)} = \frac{w^2}{M_1}
            \left(\begin{array}{ccc}
            f_0^2 x^2 & f_0 f_1 x & f_0 f_1 x  \\
            \surd & f_1^2 + \Delta f & f_1^2 - \Delta f \\
            \surd & \surd & f_1^2  + \Delta f
            \end{array} \right), \label{tbm}
    \end{eqnarray}
and
    \begin{eqnarray}
    \Delta f \equiv f_2^2 \varepsilon_2 +f_3^2 \varepsilon_3 +2 f_2 f_3 \varepsilon_x,
    \end{eqnarray}
where $x \equiv u/w,~ \varepsilon_2 \equiv M_1/M_2, ~\varepsilon_3 \equiv M_1/M_3$ and $\varepsilon_x \equiv M_1/M_x$. For simplicity, it is assumed that all $f_i$ are 1; then the mass matrix in Eq.(\ref{tbm}) has a simple pattern as follows:
    \begin{eqnarray}
    \frac{w^2}{M_1}
            \left(\begin{array}{ccc}
            x^2 & x & x \\
            \surd & 1+\Delta\varepsilon & 1-\Delta\varepsilon\\
            \surd & \surd & 1 + \Delta\varepsilon
            \end{array} \right), \label{tbm_simple}
    \end{eqnarray}
where $\Delta\varepsilon \equiv \varepsilon_2 + \varepsilon_3 + 2\varepsilon_x$.
The above matrix has a vanishing determinant, implying that one mass should be zero. The ratio of non-zero masses $m_2/m_3$ is $(2+x^2)/2(\varepsilon_2+\varepsilon_3+2\varepsilon_x)$. The type of mass hierarchy depends on the relative sizes of $u$ and $w$ and those of $M_1$ and $M_{2,3}$.
If $x=1$, furthermore, the matrix in Eq.(\ref{tbm_simple}) is exactly of the form that results from the tri-bi-maximal mixing: $U_{TBM}\cdot\mathrm{Diag}(0,m_2,m_3)\cdot U_{TBM}^T$. Even when the Yukawa coupling constants, $f_i$, are allowed to be different from each other, $\theta_{13}$ and one of the masses vanish. Only the mass ratio $m_2/m_3$ is shifted to $(2 f_1^2+f_0^2 x^2)/2(f_2^2 \varepsilon_2+f_3^2 \varepsilon_3+2f_2f_3\varepsilon_x)$. The specific pattern described above was discussed further in a previous work \cite{Park:2011zt}.

The PMNS matrix $U_{PMNS}$ is obtained by
\begin{widetext}
    \begin{eqnarray}
    U_l^\dagger U_{TBM} =
        \left(
        \begin{array}{ccc}
        \cos\theta_l & 0 & -\sin\theta_l e^{i\delta_l} \\
        0 & 1 & 0 \\
        \sin\theta_l e^{-i\delta_l} & 0 & \cos\theta_l
        \end{array} \right)
            \left(
            \begin{array}{ccc}
            \sqrt{\frac{2}{3}} & \frac{1}{\sqrt{3}} & 0 \\
            -\frac{1}{\sqrt{6}} & \frac{1}{\sqrt{3}} & \frac{1}{\sqrt{2}} \\
            \frac{1}{\sqrt{6}} & -\frac{1}{\sqrt{3}} & \frac{1}{\sqrt{2}}
        \end{array} \right), \label{pmnsProduct}
    \end{eqnarray}
where $U_l$ is the transformation of $M_l$ in Eq.(\ref{MassLepton}) that imbeds $R(\theta_l,\delta_l)$ in Eq.(\ref{TansLepton}) into the 1-3 block.
On the other hand, if $U_{PMNS}$ is expressed in the standard parametrization,
    \begin{eqnarray}
	U_{PMNS}=\left(\begin{matrix}	
		c_{12}c_{13} & c_{13}s_{12} & s_{13}e^{-i\delta} \\
		-c_{23}s_{12}-c_{12}s_{13}s_{23}e^{i\delta} &
		c_{12}c_{23}-s_{12}s_{13}s_{23}e^{i\delta} &
		c_{13} s_{23} \\
		s_{23}s_{12}-c_{12} c_{23} s_{13}e^{i\delta} &
		-c_{12} s_{23} -c_{23}s_{12} s_{13}e^{i\delta} &
		c_{13} c_{23},
    	\end{matrix}\right),\label{standard}
    \end{eqnarray}
\end{widetext}
where $c_{ij}$ and $s_{ij}$ are $\cos\theta_{ij}$ and $\sin\theta_{ij}$, respectively. From the comparison of two expressions in Eq.(\ref{pmnsProduct}) and Eq.(\ref{standard}), the three angles in PMNS matrix are obtained by the following simple relations,
    \begin{eqnarray}
    && \sin\theta_{13} = \frac{1}{\sqrt{2}}\sin\theta_l  \label{pmns}\\
    && \sin\theta_{23} = \frac{1}{\sqrt{2-\sin^2\theta_l}} \\
    && \sin\theta_{12} = \frac{\cos\theta_l - \sin\theta_l}{\sqrt{3-\frac{3}{2}\sin^2\theta_l}},
    \end{eqnarray}
where $\delta_l=\pi$ in Eq.(\ref{pmnsProduct}) and $\delta=0$ in Eq.(\ref{standard}). The prediction of the model can be estimated with respect to the $\tan2\theta_l$ in Eq.(\ref{theta}). The $\sin\theta_l$ in Eq.(\ref{pmns}) matched to $0.292<\tan2\theta_l<0.813$ can predict the range of $\theta_{13}$ measured in recent neutrino oscillation experiments \cite{Abe:2011sj,Adamson:2011qu,Abe:2011fz,An:2012eh,Ahn:2012nd}. Another predicted result is that $\theta_{23}$ belongs to the second octant. However, the range of $\theta_{12}$ predicted from the given $\theta_l$ is barely overlapped with the $3\sigma$ range of $\theta_{12}$ in the global analysis. A narrow range of $\tan2\theta_l$, $0.292<\tan2\theta_l<0.298$, can generate simultaneously the angles $0.10<\sin\theta_{13}<0.11$ and $0.51<\sin\theta_{12}<0.53$ marginally allowed in experiment, of which the eligibility is about to be tested by a higher precision data of the current oscillation experiments \cite{Abe:2011sj,Adamson:2011qu,Abe:2011fz,An:2012eh,Ahn:2012nd}.

\subsection{Neutrino model for a sizable $\theta_{13}$ }

When the $\tan 2\theta_l$ is very small as considered in Eq.(\ref{diagmm}), i.e., when $U_l$ is almost the unit, the PMNS matrix can be simply the transformation of neutrino mass matrix. The two righthanded neutrinos are in a doublet so they are not distinguishable in terms of flavor symmetry. The Majorana neutrinos $n_2$ and $n_3$ are tied into $N_{II}$, a two-dimensional representation of $\mathbb{S}_3$, so that $N_{II}=(n_2, n_3)$. The scalar particles and $n_1$ do not change their charges, and all non-SM particles have $\mathbb{Z}_2$-odd quantum number. The group representations of particles are summarized as follows:
    \begin{eqnarray}
    \begin{array}{lcl}
        (\mathbf{1},-1)_F & : & n_1 \nonumber \\
        (\mathbf{2},-1)_F & : & N_{II} ~(n_2, n_3).
        \end{array}
    \end{eqnarray}
The Higgs potential is the same as the one for Model A in Eq.(\ref{potentialNH}). The Yukawa interactions are
    \begin{eqnarray}
    -\mathcal{L}_\mathrm{ext} & = & g_0 \Sigma \overline{N}_{II} l_e + g_1 h \overline{N}_{II} L_\alpha + g_2 \Sigma \overline{n}_1 L_\alpha \nonumber \\
    & + & g_3 \Sigma \overline{N}_{II} L_\alpha. \label{extended2}
    \end{eqnarray}
The couplings of $\Sigma, ~h, ~n_i$ and $L_\alpha$ can be rephrased as $g_0 \Sigma \overline{N}_{II} l_e = g_0  (\overline{n}_2 \sigma_1 l_e + \overline{n}_3 \sigma_2 l_e)$, $g_1 h \overline{N}_{II} L_\alpha = g_1 (\overline{n}_2 h l_\mu + \overline{n}_3 h l_\tau)$, $g_2 \Sigma \overline{N}_{II} L_\alpha = g_2 \overline{n}_1 (\sigma_1 l_\mu + \sigma_2 l_\tau)$, and $g_3 \Sigma \overline{N}_{II} L_\alpha = g_3 \{ \overline{n}_2 (\sigma_2 l_\tau - \sigma_1 l_\mu) + \overline{n}_3 (\sigma_1 l_\tau + \sigma_1 l_\tau)\}$. The Majorana masses of the righthanded neutrinos are given by
    \begin{eqnarray}
        & & \frac{1}{2} M_1 n_1 n_1 + \frac{1}{2} M_2 N_{II} N_{II} \nonumber \\
        &=& \frac{1}{2} M_1 n_1 n_1 + \frac{1}{2} M_2 \left( n_2 n_2 + n_3 n_3 \right),
    \end{eqnarray}
where $N_{II} N_{II}$ becomes $n_2 n_2 + n_3 n_3$ according to the product rule in Eq.(\ref{prod3}), and breaking of $\mathbb{S}_3$ gives rise to a common mass $M_2$ for $n_2$ and $n_3$. For the same reason, the neutrinos have common Yukawa coupling constants $g_0, g_1,$ and $g_3$. For very heavy Majorana neutrinos, any two Yukawa couplings can link to each other, as shown in Fig.1.

The interactions in Figure 1 can be considered as non-renomalizable 5 dimensional couplings   obtained by integrating out the internal fermions $n_i$. Then the effective Lagrangian is given by
\begin{widetext}
    \begin{eqnarray}
    -\mathcal{L}_{eff}
    & = & \frac{g_0^2}{M_2} (\sigma_1 \sigma_1 l_e l_e + \sigma_2 \sigma_2 l_e l_e )
        +  \frac{g_0g_1}{M_2}(h \sigma_1 l_e l_\mu + h\sigma_2 l_e l_\tau ) \\
    & + & \frac{g_0g_3}{M_2}\{ -\sigma_1 \sigma_1 l_e l_\mu + \sigma_2 \sigma_2 l_e l_\mu
        + 2\sigma_1 \sigma_2 l_e l_\tau \} \\
    & + & \frac{g_2^2}{M_1} ( \sigma_1 l_\mu + \sigma_2 l_\tau )^2
        + \frac{g_1^2}{M_2}(h h l_\mu l_\mu + h h l_\tau l_\tau ) \\
    & + & \frac{g_1g_3}{M_2}\{ -h \sigma_1 l_\mu l_\mu + h \sigma_2 l_\tau l_\tau
        + h \sigma_1 l_\mu l_\tau + h \sigma_2 l_\tau l_\mu \} \\
    & + & \frac{g_3^2}{M_2} \{ ( \sigma_2 l_\tau - \sigma_1 l_\mu )^2
        + ( \sigma_1 l_\tau + \sigma_2 l_\mu )^2\},
    \label{effective2}
    \end{eqnarray}
where $M_i$ are the heavy masses of singlet Majorana neutrinos $n_i$ for $i=1-3$. When the scalar fields obtain vevs by spontaneous breaking of $\mathbb{S}_3$ symmetry as shown in Eq.(\ref{first_der})-Eq.(\ref{second_der}), the above 5-dimensional interactions reduce to low-energy effective mass terms of light neutrinos $M^{(\nu)}_{ij}\nu_i\nu_j$. The matrix is
    \begin{eqnarray}
    M^{(\nu)} &=& \frac{w^2}{M_1} \left(\begin{array}{ccc}
            2 g_0^2 \varepsilon
            & g_0 g_1 x\varepsilon
            & g_0 g_1 x\varepsilon + 2 g_0 g_3 \varepsilon  \\
            \surd
            & g_2^2 + g_1^2x^2\varepsilon - g_1 g_3 x \varepsilon + 2 g_3^2 \varepsilon
            & g_2^2 + g_1 g_3 x \varepsilon  \\
            \surd
            & \surd
            & g_2^2 + g_1^2x^2\varepsilon + g_1 g_3 x \varepsilon + 2 g_3^2 \varepsilon
            \end{array} \right),  \label{t2kmass}
    \end{eqnarray}
\end{widetext}
where $x \equiv u/w,$ and $\varepsilon \equiv M_1/M_2$. For the simplest analysis, it is assumed that all $g_i$'s are one except $g_0$. The $g_0$ is smaller than 1 to make $m_1$ smaller than $m_2$.
\begin{figure}
\resizebox{80mm}{!}{\includegraphics[width=0.75\textwidth]{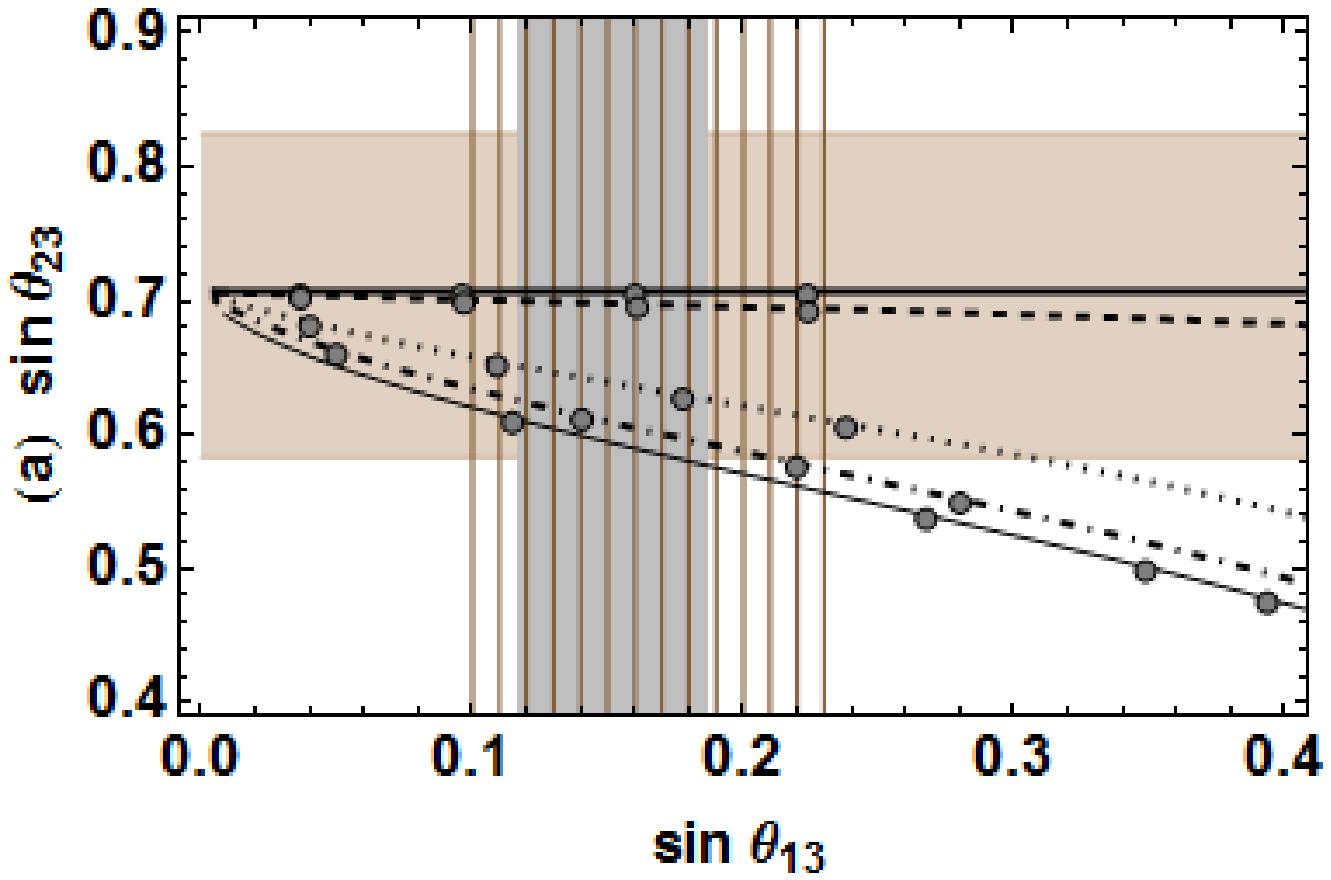}}
\resizebox{80mm}{!}{\includegraphics[width=0.75\textwidth]{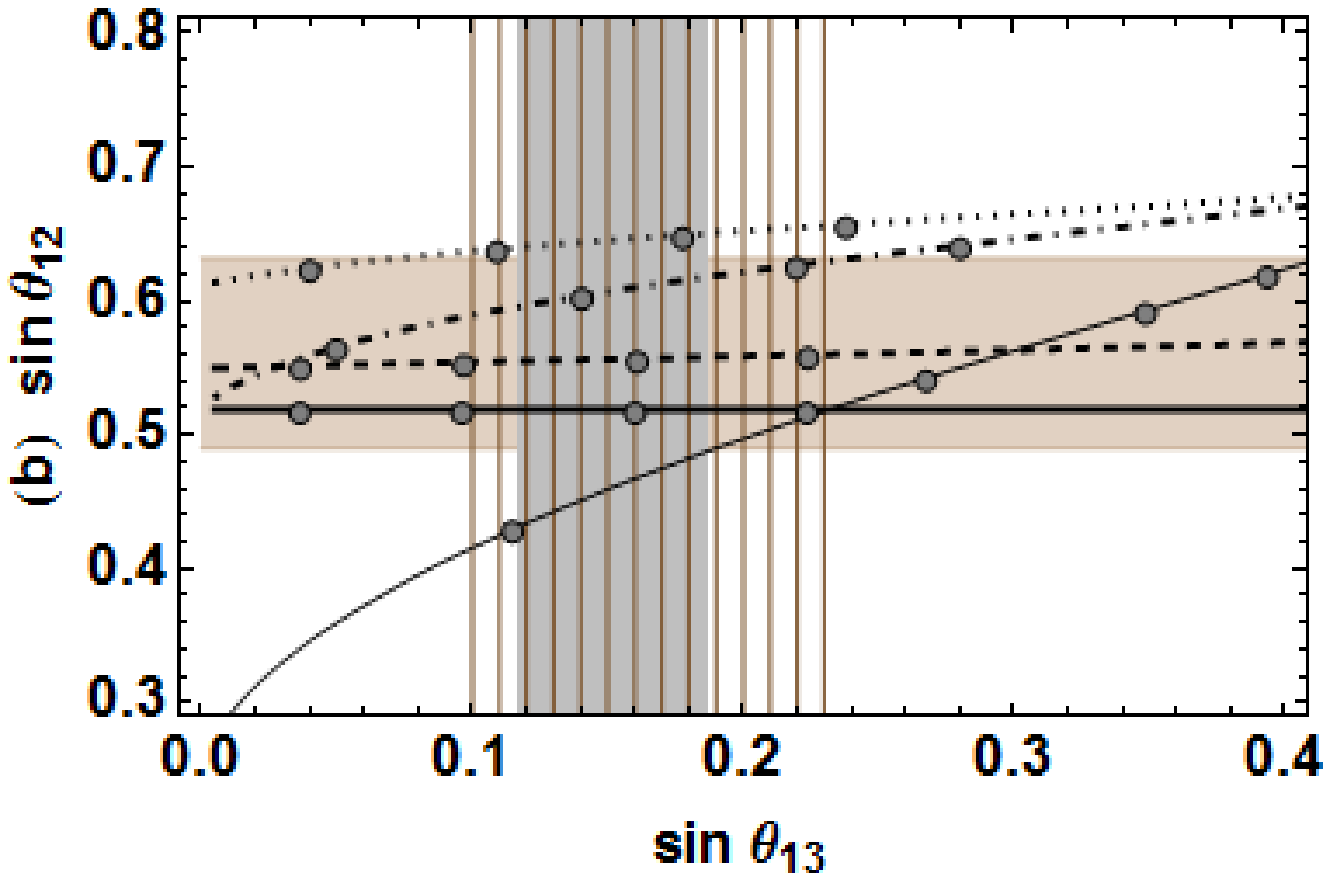}}
\resizebox{80mm}{!}{\includegraphics[width=0.75\textwidth]{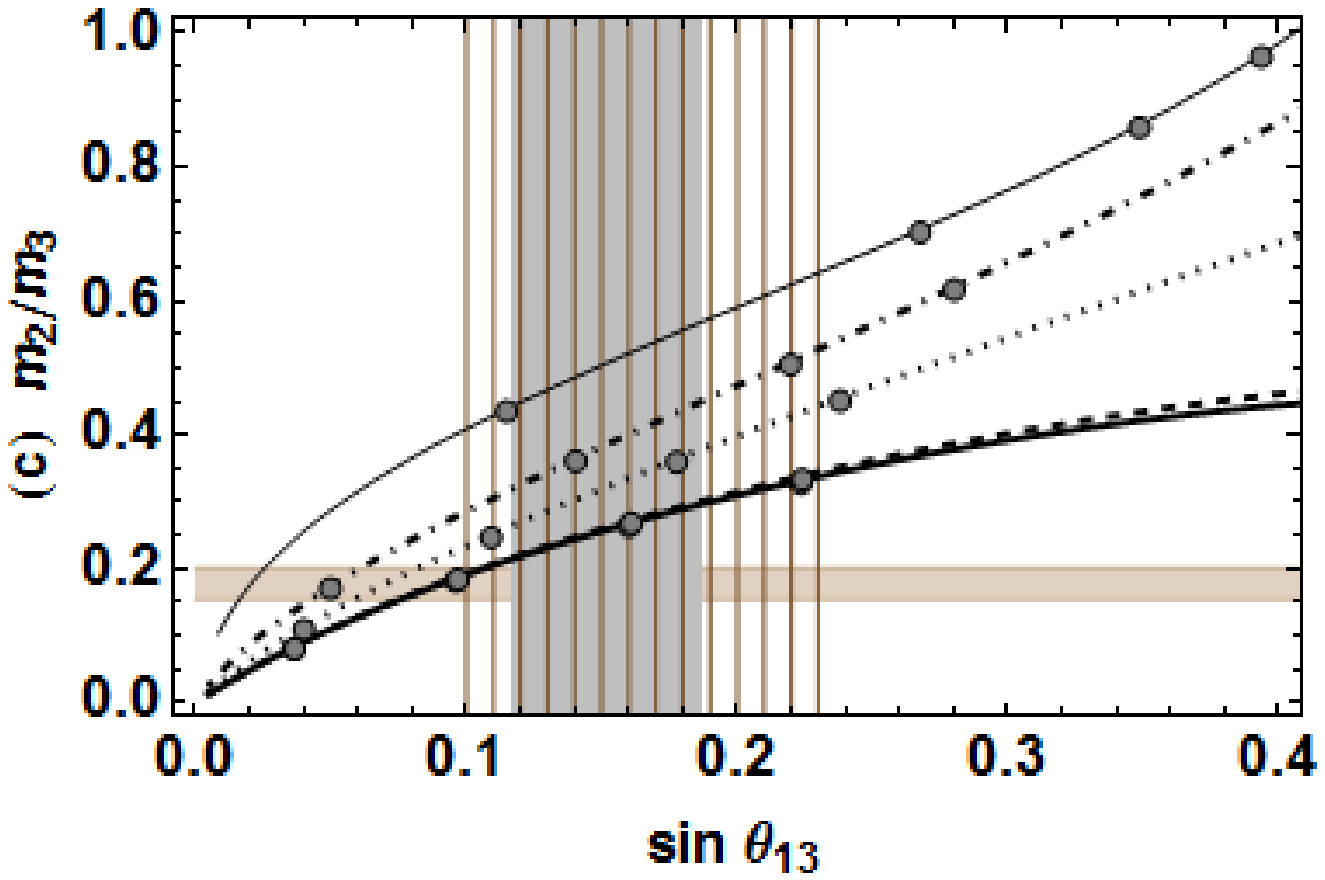}}
\caption{\label{fig:rge_msqr} Predictions by Model B. The brown shadow in each figure indicates the bound at the 90\% CL for the angle and the mass ratio: $0.583<\sin\theta_{23}<0.825$, $0.490<\sin\theta_{12}<0.632$, and $0.154<\sqrt{\Delta m_{21}^2/ \Delta m_{32}^2}<0.201$. As for $\sin\theta_{13}$, the brown striped region indicates the RENO result: $0.100<\sin\theta_{13}<0.237$, and the gray shadowed region indicates Daya Bay result: $0.117<\sin\theta_{13}<0.187$. The thick solid, dashed, dotted, dot-dashed, and thin solid lines indicate $\langle h \rangle/\langle \Sigma \rangle \equiv x=0.0001, ~0.1, ~1.0, ~2.0$ and $~5.0$ in Eq.(\ref{t2kmass}). The four small circles on each curve indicate $M_1/M_2 \equiv\varepsilon=0.07, ~0.17, ~0.27$ and $~0.37$ in Eq.(\ref{t2kmass}).}
\end{figure}

A rather tedious estimation of masses and mixing angles from the above mass matrix is presented using pictorial analysis. Fig.2 presents the possible parametric plots that can arise from Eq.(\ref{t2kmass}) in spaces of $\sin\theta_{13}\mathrm{vs.}\sin\theta_{23}$, $\sin\theta_{13}\mathrm{vs.}\sin\theta_{12}$, and $\sin\theta_{13}\mathrm{vs.}m_2/m_3$, respectively. The value of $g_0$ is chosen as 0.7 in the following example. Five curves in each figure represent the choices of $x\equiv u/w$ as 0.0001, 0.1, 1.0, 2.0, and 5.0 by thick solid, dashed, dotted, dot-dashed, and thin solid lines, respectively. Four circles on each curve represent the choices of $\varepsilon\equiv M_1/M_2$ as 0.07, 0.17, 0.27, and 0.37 as $\sin\theta_{13}$ increases. The horizontal shadow in each figure represents the allowed region at the 90\% CL for the physical parameters according to the current experimental results: $0.583<\sin\theta_{23}<0.825$, $0.590<\sin\theta_{12}<0.632$, and $0.154<\sqrt{\Delta m_{21}^2/ \Delta m_{32}^2}<0.201$. As for the mass ratio, the region above 0.201 is not ruled out since $m_2^2/m_3^2$ can be larger than $\Delta m_{21}^2/ \Delta m_{32}^2$. For $\sin\theta_{13}$, two bounded regions are presented. One is $0.100<\sin\theta_{13}<0.237$ from RENO \cite{Ahn:2012nd}, which is indicated by brown stripes. The other is $0.117<\sin\theta_{13}<0.187$ from Daya Bay \cite{An:2012eh}, which is indicated by gray shadows.

The bundle of curves in Figure 2 shows the area that the model can cover as $M_1/M_2$ increases. Different curves in a figure come from the relative ratio of the vevs of non-SM Higgs fields. The construction of mass matrix does not derive coefficients of elements, leaving them free parameters, when the symmetry builds up a pattern of a mass matrix based on the charge assignments of a flavor symmetry. Here, we examine the prediction from the model, while the effect of Yukawa couplings, $f_i$ or $g_i$, is suppressed by setting them one. Whatever the value of $M_1/M_2$ is and whatever the ratio $\langle h \rangle/\langle \Sigma \rangle$ is, there is some area in $\sin\theta_{23}$, which is larger than 0.707, excluded by the prediction.

As for $m_2/m_3$, likewise, the mass type of degeneracy or quasi-degeneracy is ruled out. The thin solid line in each figure describes the fit for $\langle h \rangle/\langle \Sigma \rangle=5$, and figures 2(b) and 2(c) show that the resulted curves miss the allowed range, if $\langle h \rangle/\langle \Sigma \rangle >5$. So the mass ratio range is confined within $m_2/m_3 < 0.6$ at most.

On the other hand, the ranges of $M_1/M_2$ and $\langle h \rangle/\langle \Sigma \rangle$ are also trimmed off by the experimental bounds of mixing angles. For instance, if $\langle h \rangle/\langle \Sigma \rangle < 0.1$, any value of $M_1/M_2$ smaller than 0.17 and that larger than 0.40 are excluded by RENO bound on $\theta_{13}$.

\section{Conclusion}

The recent measurements of $\sin^2(2\theta_{13})$ motivate an idea which is that non-zero $\sin\theta_{13}$ is generated by a mechanism based on the symmetrical background rather than being a perturbation effect from TBM with $\sin\theta_{13}=0$. Two lepton models were introduced in terms of $\mathbb{S}_3\otimes\mathbb{Z}_2$ flavor symmetry. One provides TBM from $U_\nu$ and non-zero $\theta_{13}$ from $U_l$ to the PMNS matrix, while the other provides all leading orders of mixing angles from the neutrino mixing matrix, $U_\nu$. The difference between two models is caused by the only difference between the flavor charge assignments for righthanded neutrinos. Other group theoretical properties are all the same for both models.

The $\mathbb{Z}_2$ symmetry splits the field contents into the particle fields beyond the SM and the fields in the SM, whether the charge is -1 or +1. Since the SM fields are $\mathbb{Z}_2$-even, the Yukawa couplings of the SM fermions are protected from the coupling with a $\mathbb{Z}_2$-odd scalar field. On the other hand, a $\mathbb{Z}_2$-odd righthanded neutrino makes a vertex with a $\mathbb{Z}_2$-odd scalar field. If a righthanded neutrino as an internal line is integrated out and the effective 5-dimensional coupling is suppressed by the mass scale of the righthanded neutrino, the Majorana masses of lefthanded neutrinos become then light. The $\mathbb{Z}_2$-even scalar Higgs fields, $H$ and $\Phi$, in Yukawa couplings contribute to the masses of charged leptons, and $\mathbb{Z}_2$-odd scalar Higgs fields, $h$ and $\Sigma$, in effective 5-dimensional couplings contribute to the masses of neutrinos.

Depending on the flavor charges of $n_2$ and $n_3$ among three generations, the type of neutrino mixing was determined. When they belong to separate $(\mathbf{1'},-1)_F$ representations, the model gave rise to the exact $\sin\theta_{13}^\nu=0$, as shown in Eq.(\ref{tbm_simple}). When they belong to a single $(\mathbf{2},-1)_F$ representation, the model gave rise to $\sin\theta_{13}^\nu\neq0$ unless $M_1/M_2=0$. The prediction of the model, neglecting the contributions from most $g_i$, was studied in Fig.2. The results obtained for various ranges of the relative ratio, $\frac{M_1}{M_2}$, of Majorana masses and for those of the relative scales of vevs of non-SM Higgs, $\frac{\langle h \rangle}{\langle \Sigma \rangle}$, rule out the area of $\sin\theta_{23}$ larger than 0.707 and the mass pattern of (quasi-) degeneracy. Thus, the survival of model B can be determined, depending on whether $\theta_{23}<\pi/4$ and whether the mass type is hierarchical.

\begin{acknowledgments}
This work was supported by the Basic Science Research program through NRF(2011-0014686).
\end{acknowledgments}

\appendix

\section{Higgs Potential}
The contents of Higgs scalar particles and their representations under $\mathbb{S}_3\otimes\mathbb{Z}_2$ are
    \begin{eqnarray}\begin{array}{lll}
        (\mathbf{1},1)_F & : & H \\
        (\mathbf{2},1)_F & : & \Phi~(\varphi_1, \varphi_2) \\
        (\mathbf{1},-1)_F & : & h \\
        (\mathbf{2},-1)_F & : & \Sigma~(\sigma_1, \sigma_2), \label{4higgs}
    \end{array}
    \end{eqnarray}
which commonly belong to $(\mathbf{2},1/2)_G$ under $SU(2)\times U(1)$ gauge group.
The full invariant Higgs potential can be organized into three parts as follows:
    \begin{eqnarray}
    V=V_e(H,\Phi)+V_o(h,\Sigma)+V_\chi(H,\Phi;h,\Sigma), \label{3potential}
    \end{eqnarray}
where $V_e$ and $V_o$ are the interactions of only $\mathbb{Z}_2$-even particles and those of only $\mathbb{Z}_2$-odd particles, respectively, while $V_\chi$ is the cross interactions of $\mathbb{Z}_2$-even and $\mathbb{Z}_2$-odd particles. Each contribution to the potential $V$ is given as;
\begin{widetext}
    \begin{eqnarray}
        && V_e(H,\Phi) = \label{potentialNH}
            m_H^2H^\dagger H + \frac{1}{2}\eta(H^\dagger H)^2 + m_\varphi^2 \Phi^\dagger\Phi +\frac{1}{2}\Lambda(\Phi^\dagger\Phi)_r^2 \\
        && \hspace{10pt}+ ~\lambda(\Phi^\dagger\Phi)_1(H^\dagger H)_1 +
            \lambda'(\Phi^\dagger H)_2(H^\dagger\Phi)_2+\lambda''\{(\Phi^\dagger H)_2^2 + h.c.\} +\kappa\{(\Phi^\dagger\Phi)_2(\Phi^\dagger H)_2 + h.c.\}, \nonumber \\
            \nonumber \\
        && V_o(h,\Sigma)  =  \label{potential_sigma}
            m_h^2h^\dagger h + \frac{1}{2}\lambda_h(h^\dagger h)^2
            + m_s^2 \Sigma^\dagger\Sigma +
            \frac{1}{2}\Lambda_s(\Sigma^\dagger\Sigma)_r^2 \\
        && \hspace{10pt}+ ~\lambda_s(\Sigma^\dagger\Sigma)_1(h^\dagger h)_1 +
            \lambda_s'(\Sigma^\dagger h)_2(h^\dagger\Sigma)_2 +
            \lambda_s''\{(\Sigma^\dagger h)_2^2+\mathrm{h.c.}\} + \kappa_s\{(\Sigma^\dagger\Sigma)_2(\Sigma^\dagger h)_2+\mathrm{h.c.}\}.\nonumber \\
            \nonumber \\
        && V_\chi(H,\Phi;h,\Sigma) =  \label{potential_ext}
        \chi(H^\dagger H)_1(h^\dagger h)_1+\chi'(H^\dagger h)_1(h^\dagger H)_1+
            \chi''\{(H^\dagger h)_1^2+\mathrm{h.c.}\} \\
        && \hspace{10pt}+ ~\lambda_\chi(\Phi^\dagger \Phi)_1(h^\dagger h)_1 +
            \lambda_\chi'(\Phi^\dagger h)_2(h^\dagger \Phi)_2+
            \lambda_\chi''\{(\Phi^\dagger h)_2^2+\mathrm{h.c.}\} \nonumber \\
        && \hspace{10pt}+ ~\eta_\chi(\Sigma^\dagger\Sigma)_1(H^\dagger H)_1 +
            \eta_\chi'(\Sigma^\dagger H)_2(H^\dagger \Sigma)_2+
            \eta_\chi''\{(\Sigma^\dagger H)_2^2+\mathrm{h.c.}\} \nonumber \\
        && \hspace{10pt}+ ~\gamma\{(H^\dagger h)_1(\Sigma^\dagger\Phi)_1+\mathrm{h.c.}\} +
            \gamma'\{(H^\dagger \Sigma)_2(h^\dagger\Phi)_2+\mathrm{h.c.}\} \nonumber \\
        && \hspace{10pt}+ ~\Gamma_\chi(\Phi^\dagger \Phi)_r(\Sigma^\dagger\Sigma)_r +
            \Gamma_\chi'(\Phi^\dagger\Sigma)_r(\Sigma^\dagger\Phi)_r+
            \Gamma_\chi''\{(\Phi^\dagger \Sigma)_r^2+\mathrm{h.c.}\}.
            \nonumber
    \end{eqnarray}
\end{widetext}
The subscript `1' or `2' in each term indicates that the product of two fields belongs to the representation $\mathbf{1}$ or $\mathbf{2}$ in $\mathbb{S}_3$. Each term with a subscript `$r$' consists of three types of products, $\mathbf{1}, ~\mathbf{1'}$ and $\mathbf{2}$ representations as in Eq.(\ref{3reps}).

According to the product rules in Eqs. (\ref{prod1}) - (\ref{prod3}),
$(\Phi^\dagger\Phi)_1=|\varphi_1|^2+|\varphi_2|^2,~(\Phi^\dagger\Phi)_{1'}=\varphi_1^*\varphi_2-\varphi_2^*\varphi_1$,
and $(\Phi^\dagger\Phi)_2=(|\varphi_2|^2-|\varphi_1|^2 ~ ~ \varphi_1^*\varphi_2+\varphi_2^*\varphi_1)^T$.
The  Higgs potential in Eq.(\ref{potentialNH}) can be rephrased in terms of component fields $\{\varphi_i, \varphi_i^\dagger\}$ with $i=1$ and 2, and $\{H, H^\dagger\}$:
\begin{widetext}
    \begin{eqnarray}
    && V_e~(H,~H^\dagger,~\varphi_i,\varphi_i^\dagger) = m_H^2|H|^2 + \frac{1}{2}\eta|H|^4 \\
    && \hspace{10pt}+ \left(m_\varphi^2 + (\lambda+\lambda')|H|^2 \right)\sum_{i}|\varphi_i|^2 + \lambda''\{ H^2(\varphi_1^{*2}+ \varphi_2^{*2}) + \mathrm{h.c.}\} \nonumber \\
    && \hspace{10pt}+ \frac{1}{2}(\lambda_a+\lambda_c)\sum_{i}|\varphi_i|^4
    + (\lambda_a+\lambda_b)|\varphi_1|^2|\varphi_2|^2 \frac{1}{2}(\lambda_c-\lambda_b)(\varphi_1^{*2}\varphi_2^2 + \varphi_2^{*2}\varphi_1^2) \nonumber \\
    && \hspace{10pt}+ \kappa\{ H (2|\varphi_2|^2\varphi_1^*+(\varphi_2^*)^2\varphi_1-|\varphi_1|^2\varphi_1^*) + \mathrm{h.c.}\}. \nonumber
    \end{eqnarray}
When the Higgs particles obtain their real vacuum expectation values such that $\langle H\rangle=\langle H^\dagger\rangle=v$,  $\langle\varphi_1\rangle=v_1$, and $\langle\varphi_2\rangle=v_2$, the potential can be expressed as follows.
    \begin{eqnarray}
    && V_e~(v,v_1,v_2) ~= ~m_H^2v^2 + m_\varphi^2(v_1^2+v_2^2) + \frac{1}{2}\eta v^4 \label{potential_vev} \\
    && \hspace{10pt}+ ~\frac{1}{2}\Lambda_a(v_1^2+v_2^2)^2 +\Lambda_b v^2(v_1^2+v_2^2)
    + 2\kappa v(3v_2^2v_1-v_1^3), \nonumber
    \end{eqnarray}
\end{widetext}
where $\Lambda_a=\lambda_a+\lambda_c$, and $\Lambda_b=\lambda+\lambda'+2\lambda''$.

Following the same steps as in Eq.(\ref{potentialNH}) - Eq.(\ref{potential_vev}), the potentials, $V_o$ and $V_\chi$, in terms of vevs, $\langle h\rangle=u$ and $(\langle \sigma_1\rangle, \langle \sigma_2 \rangle)=(w_1,w_2)$, can be expressed as follows:
\begin{widetext}
    \begin{eqnarray}
    && V_o~(u,w_1,w_2) ~= ~m_h^2u^2 + m_s^2(w_1^2+w_2^2) + \frac{1}{2}\lambda_h u^4 \label{potential_odd} \\
    && \hspace{10pt}+ ~\frac{1}{2}\Lambda_s(w_1^2+w_2^2)^2 +\Lambda_c u^2(w_1^2+w_2^2) + ~2\kappa_s u(3w_2^2w_1-w_1^3), \nonumber
    \end{eqnarray}
where $\Lambda_c \equiv \lambda_s+\lambda_s'+2\lambda_s''$.
    \begin{eqnarray}
    && V_\chi = k_1u^2v^2 + k_2u^2(v_1^2+v_2^2) +k_3v^2(w_1^2+w_2^2) +k_4 uv(v_1w_1+v_2w_2)
            \label{potential_cross} \\
    && \hspace{10pt}+ k_5v_1v_2w_1w_2
        +k_5'(v_1^2+v_2^2)(w_1^2+w_2^2)+ k_5''(v_2^2-v_1^2)(w_2^2-w_1^2) +k_5'''(v_1^2w_1^2+v_2^2w_2^2), \nonumber
    \end{eqnarray}
\end{widetext}
where $k_1=\chi+\chi'+2\chi'', ~k_2=\lambda_\chi+\lambda_\chi'+2\lambda_\chi'', ~k_3=\eta_\chi+\eta_\chi'+2\eta_\chi'',$ and $k_4=2(\gamma+\gamma')$. The $k_5...k_5'''$ are rather complicated polynomials of $\Gamma_\chi,\Gamma'_\chi,$ and $\Gamma_\chi$ in Eq.(\ref{potential_ext}), such that $k_5=k_5(\Gamma_\chi, \Gamma'_\chi), ~k_5'=k_5'(\Gamma_\chi, \Gamma''_\chi), ~k_5''=k_5''(\Gamma_\chi),$ and $k_5'''=k_5'''(\Gamma'_\chi)$.
Their details are not necessary for the following examination of the minimal condition.
The first derivatives of the full potential given in Eq.(\ref{3potential}) are
\begin{widetext}
    \begin{eqnarray}
    \frac{\partial V}{\partial v} &=& 2v \{ K(m_H^2, u^2, v_i^2, w_i^2) +\eta v^2 \}
        + 2\kappa (3v_2^2v_1-v_1^3) + k_4u(v_1w_1+v_2w_2) \label{6min}\\
    \frac{\partial V}{\partial u} &=& 2u \{ K(m_h^2, v^2, v_i^2, w_i^2) +\lambda_h u^2 \}
        + 2\kappa_s (3w_2^2w_1-w_1^3) + k_4 v (v_1w_1+v_2w_2) \nonumber \\
    \frac{\partial V}{\partial v_1} &=& 2v_1 \{ K(m_\varphi^2, u^2, v^2, w_i^2)
        +\Lambda_a (v_1^2+v_2^2) \} + 6\kappa v(v_2^2-v_1^2) + k_4uvw_1 + k_5v_2w_1w_2 \nonumber \\
    \frac{\partial V}{\partial v_2} &=& 2v_2 \{ K(m_\varphi^2, u^2, v^2, w_i^2)
        +\Lambda_a (v_1^2+v_2^2) \} + 12\kappa vv_1v_2 + k_4uvw_2 + k_5v_1w_1w_2 \nonumber \\
    \frac{\partial V}{\partial w_1} &=& 2w_1 \{ K(m_s^2, u^2, v^2, v_i^2)
        +\Lambda_s (w_1^2+w_2^2) \} + 6\kappa_s u (w_2^2-w_1^2) + k_4uvv_1 + k_5v_1v_2w_2\nonumber \\
    \frac{\partial V}{\partial w_2} &=& 2w_2 \{ K(m_s^2, u^2, v^2, v_i^2)
        +\Lambda_s (w_1^2+w_2^2) \} + 12\kappa_s u w_1w_2 + k_4uvv_2 + k_5v_1v_2w_1, \nonumber
    \end{eqnarray}
\end{widetext}
where each $K$ denotes the part that corresponds to the coefficients of linear terms.
The vevs, $v_1\neq0$ and $v_2=0$, can make the potential minimum, when the following conditions are satisfied.
    \begin{eqnarray}
    &&\left(\frac{\partial V}{\partial v_1}\right)_{v_2=0} = 2v_1(K+\Lambda_a v_1^2)
        -6\kappa v v_1^2 +k_4uvw_1 =0 \nonumber \\
    &&\left(\frac{\partial V}{\partial v_2}\right)_{v_2=0} = k_4uvw_2 +k_5v_1w_1w_2 =0
    \label{first_der}
    \end{eqnarray}
    \begin{eqnarray}
    &&\left(\frac{\partial^2 V}{\partial v_1\partial v_2}\right)_{v_2=0} = k_5w_1w_2 >0
    \label{second_der} \\
    &&\left(\frac{\partial^2 V}{\partial w_1\partial w_2}\right)_{v_2=0} =
    4\Lambda_sw_1w_2 +12\kappa_s uw_2 >0. \nonumber
    \end{eqnarray}
It is clear that any of $w_1$ and $w_2$ should not be zero to satisfy the above conditions. According to the symmetry of the potential under the interchange of $\sigma_1$ and $\sigma_2$, vevs can be taken as $w_1=w_2=w$. Thus, in summary, the following vevs of the fields in Eq.(\ref{4higgs}) can be adopted for the masses of leptons:
    \begin{eqnarray}\begin{array}{lll}
        \langle H \rangle & = & v\\
        \langle \Phi \rangle & = & (v_1, 0) \\
        \langle h \rangle & = & u\\
        \langle \Sigma \rangle & = & (w, w). \label{4vevs}
    \end{array}
    \end{eqnarray}
Then, the derivatives in Eqs.(\ref{6min}) reduces to the following conditions:
\begin{widetext}
    \begin{eqnarray}
    && 2v \{ m_H^2 + k_1 u^2 + 2 k_3 w^2 +\eta v^2 \}
        - 2\kappa v_1^3 + k_4u v_1w =0 \label{4min}\\
    && 2u \{ m_h^2 + k_1 v^2 + 2 k_2 v_1^2+\lambda_h u^2 \}
        + 4\kappa_s w^2 + k_4 v v_1w =0 \nonumber \\
    && 2v_1 \{ m_\varphi^2 +\Lambda_a v_1^2 + \Lambda_b v^2 +k_2 u^2 +(2k_5'+k_5''')w^2 \}
        - 6\kappa v v_1^2 + k_4uvw =0 \nonumber \\
    && 2w \{ 2m_s^2 + 4\Lambda_s w^2 + 2\Lambda_c u^2 + 2 k_3 v^2 + +(2k_5'+k_5''')v_1^2 \}
        + k_4uvv_1 =0 \nonumber
    \end{eqnarray}
\end{widetext}
According to Eq.(\ref{first_der}) and Eq.(\ref{second_der}), $k_4<0$ and $k_5>0$ are necessary. The mass matrices are examined upon the assumptions of $0.0001<u/w<5$ and $v/v_1<1$ with weak hierarchy. The assumptions do not show any conflicts with either the minimum conditions in Eq.(\ref{4min}) or the phenomenological constraints, $\mathcal{O}(m_h^2, m_\varphi^2, m_s^2)>m_H^2$, since the constraints on masses and vevs contain a sufficient number of independent parameters.

\end{document}